# Onset Temperatures for Superconducting Fluctuations in Te-annealed FeTe$_{1-x}$Se$_x$ Single Crystals: Evidence for the BCS-BEC Crossover

Yu UEZONO[1], Takumi OTSUKA[1], Shotaro HAGISAWA[1], Haruka TANIGUCHI[2], Michiaki MATSUKAWA[2], Takenori FUJII[3], and Takao WATANABE[1*]

[1]*Graduate School of Science and Technology, Hirosaki University, Hirosaki, 036-8561, Japan*
[2]*Graduate School of Engineering, Iwate University, Morioka 020-8551, Japan*
[3]*Cryogenic Research Center, University of Tokyo, Bunkyo, Tokyo 113-0032, Japan*

*E-mail: twatana@hirosaki-u.ac.jp*

Recently, the superconductors' community has witnessed an unsettled debate regarding whether iron-based superconductors, in particular FeSe and FeSe$_{1-x}$S$_x$, are in the Bardeen–Cooper–Shrieffer (BCS) – Bose–Einstein condensation (BEC) crossover regime. Nonetheless, one particular system, FeTe$_{1-x}$Se$_x$, has been less investigated in this regard owing to the screening of its intrinsic superconducting properties by the inevitable iron excess. Herein, the onset temperatures for superconducting fluctuations ($T_{scf}$) are investigated by measuring the magnetoresistance (*MR*) of Te-annealed, high-quality FeTe$_{1-x}$Se$_x$ ($x$ = 0.1, 0.2, 0.3, and 0.4) single crystals. The results reveal very high $T_{scf}$ values for these crystals. Particularly for $x$ = 0.4, $T_{scf}$ reaches approximately 40 K, which is 2.7 times larger than $T_c$. This indicates that the superconductivity of the FeTe$_{1-x}$Se$_x$ system is well within the BCS-BEC crossover regime.

## 1. Introduction

Emergent electronic properties of iron-based superconductors have been at the forefront of condensed matter physics since their discovery in 2006 [1]. Recently, the peculiar properties of FeSe and FeSe$_{1-x}$S$_x$, such as giant superconducting fluctuations observed by magnetic torque measurements [2] and gap opening at $k$ = 0 observed by quasiparticle interference (QPI) patterns [3] or angle-resolved photoemission spectroscopy (ARPES) [4], have attracted significant attention. This is mainly owing to their possible categorization in the crossover regime between the weak-coupling Bardeen–Cooper–Shrieffer (BCS) and strong-coupling Bose–Einstein condensation (BEC) limits [5]. These unusual properties are thought to originate from their superconducting gap sizes, which are comparable to Fermi energies. However, some groups refuted observing such large superconducting fluctuations with the same magnetic torque measurements [6].

Additional iron-based superconductors, such as FeTe$_{1-x}$Se$_x$, are also expected to have similar categorization in the BCS-BEC crossover regime. The strong coupling nature of superconductivity has been reported by specific heat measurements [7,8]. The characteristic flat-band dispersion near $k = 0$, which is expected in the BEC regime, has been reported by ARPES measurements [9]. However, FeTe$_{1-x}$Se$_x$ systems have been less investigated compared to FeSe and FeSe$_{1-x}$S$_x$ systems, and transport evidence for their BCS-BEC crossover remains lacking. This is primarily owing to excess Fe, which is inevitably incorporated into the crystals; this incorporation screens their intrinsic superconducting properties. To overcome the iron excess difficulty, we have developed a novel efficient annealing method, called "Te-anneal", wherein single crystals are annealed under tellurium vapor [10,11].

In this study, we conducted magneto transport measurements for Te-annealed, high-quality FeTe$_{1-x}$Se$_x$ single crystals. We estimated the strength of Cooper pairing by measuring the superconducting resistive transitions under several magnetic fields up to 9 T. We further measured the magnetoresistance (*MR*) at various fixed temperatures to estimate the onset temperatures ($T_{scf}$) for the superconducting fluctuations. Based on these results, we discussed the possibility of BCS-BEC crossover in the FeTe$_{1-x}$Se$_x$ system.

## 2. Experiment

Single crystals of Fe$_{1+y}$Te$_{1-x}$Se$_x$ ($x$ = 0.1, 0.2, 0.3, 0.4, and $y$ = 0.03) with nominal compositions were grown using the Bridgman method [10]. The obtained crystals were cleaved into smaller samples (approximately 1-mm thick) and Te-annealed for more than 400 h at 400 °C [10,11]. Magnetic susceptibility measurements showed sharp superconducting transitions ($\Delta T_c \leq 1$ K) for each crystal, indicating a complete removal of excess iron [12].

The in-plane resistivity $\rho_{ab}$ was measured using a physical property measurement system (PPMS) (Quantum Design) with the applied fields $B$ parallel to the *c*-axis. The value of *MR* was obtained by averaging the data set at positive and negative fields, that is, $MR(B) = [\{\rho_{ab}(B) + \rho_{ab}(-B)\}/2 - \rho_{ab}(0)]/\rho_{ab}(0)$, which can eliminate the Hall component owing to the misalignment of contacts.

## 3. Results

### 3.1 Upper critical fields in the orbital-limit ($B_{c2}(0)$)

Figures 1(a) - (d) show the in-plane resistive transition curves under several magnetic fields of the Te-annealed FeTe$_{1-x}$Se$_x$ single crystals for $x$ = 0.4, 0.3, 0.2, and 0.1, respectively. For $x$ = 0.4 and 0.3, the superconducting transition width increases with increasing magnetic field, which is similar to the high-$T_c$ cuprates [13]. This implies that the interaction mediating Cooper pairs is strong (i.e. coherence length $\xi_{ab}$ is short) and the effects of superconducting fluctuations are enhanced for these crystals [14]. This behavior reproduces previous results for bulk [15] and thin film [16] samples. In contrast, the transition curves for $x$ = 0.2 and 0.1 (which have been less reported) are found to approach parallel-shifts. The reason for this will be discussed in a separate manuscript.

To evaluate the strength of Cooper pairing, the upper critical field in the orbital limit at zero temperature, $B_{c2}(0)$, was estimated using the Werthamer–Helfand–Holhenberg (WHH) formula [17]. Although FeTe$_{1-x}$Se$_x$ is known to be a multi-band system and its upper critical field may be limited by the Pauli paramagnetic effect [15], the orbital-limit $B_{c2}(0)$ estimated by the WHH formula, considering only a single band, provides a measure for the pairing strength [16]. Here, $B_{c2}$ was determined by half of the normal-state resistivity $\rho_n$ (Figs. 1(a) - (d)). The obtained $B_{c2}$ are plotted as a function of temperature in Fig. 1(e).

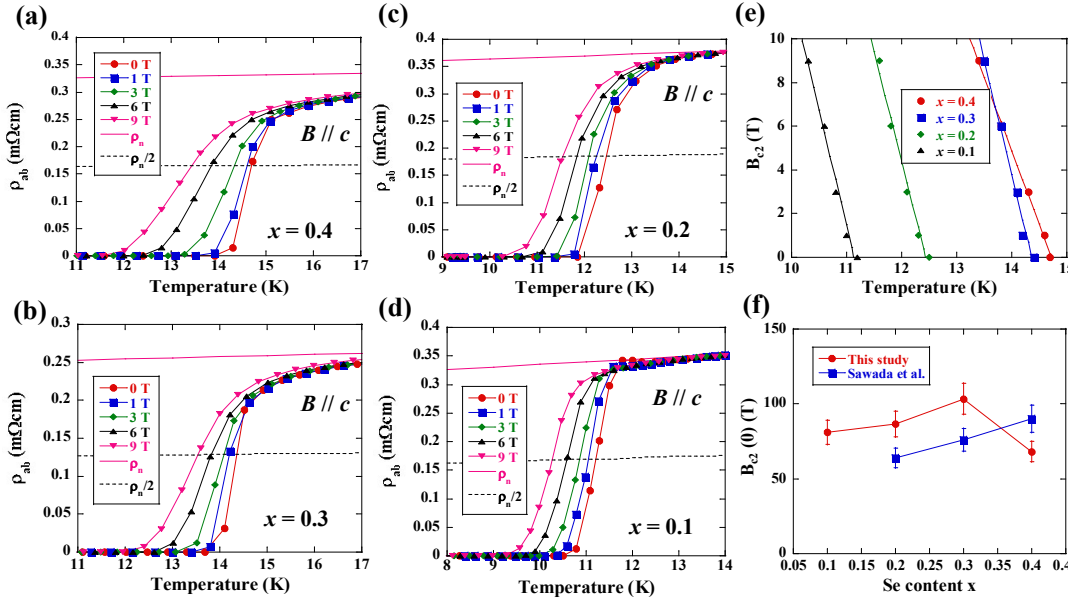

**Fig. 1.** (Color online) Temperature dependence of the in-plane resistivity $\rho_{ab}$ under magnetic fields $B//c$ of up to 9 T for the Te-annealed FeTe$_{1-x}$Se$_x$ single crystals with (a) $x = 0.4$, (b) $x = 0.3$, (c) $x = 0.2$, and (d) $x = 0.1$. Normal state resistivity $\rho_n$ was defined as the linear fit to $\rho_{ab}$ in the temperature range it changes linearly above $T_{scf}$. (e) Temperature dependence of $B_{c2}$ of the Te-annealed FeTe$_{1-x}$Se$_x$ single crystals with $x = 0.4, 0.3, 0.2$, and $0.1$. Solid lines are linear fits to the data (markers). (f) Se content $x$ dependence of $B_{c2}$ at 0 K, $B_{c2}(0)$. The red closed circles represent the data for the Te-annealed FeTe$_{1-x}$Se$_x$ single crystals, whereas blue solid squares represent those obtained for FeTe$_{1-x}$Se$_x$ thin films [16].

The temperature dependence near $T_c$ was observed to be almost linear, and the slope $-dB_{c2}/dT|_{T=T_c}$ was as high as approximately 10 T/K. From these results, $B_{c2}(0) \left(= -0.69 \cdot dB_{c2}/dT|_{T=T_c} \cdot T_c\right)$ was estimated and plotted in Fig. 1(f). The obtained $B_{c2}(0)$ was 60–100 T, which agrees with previously reported data [16]. These high $B_{c2}(0)$ values indicate that the interactions mediating Cooper pairs in this system are very strong.

*3.2 Onset temperatures for the superconducting fluctuations ($T_{scf}$)*

Strong coupling is expected to enhance $T_{scf}$ [2]. Figures 2(a) - (d) show the magnetic field dependence of the *MR* at different temperatures for the Te-annealed FeTe$_{1-x}$Se$_x$ single crystals (i.e., $x = 0.4, 0.3, 0.2$, and $0.1$). For $x = 0.4$, a negative *MR* was observed with decreasing temperature from 70 to 50 K, which likely originates from the suppression of the pseudogap by applying magnetic fields [12]. However, an additional positive *MR* was observed for temperatures below 40 K. At 40 and 36 K, the addition tends to saturate at higher fields, and the *MR* seems to recover a normal state $B^2$-dependence (although it is negative) above $B_{c2}$. As temperature decreases, it becomes difficult to ascertain that the normal state is restored, since $B_{c2}$ becomes comparable (or higher) to the highest available field. Nevertheless, the tendency towards saturation at higher fields is still evident at 32 and 30 K. Based on these observations, we attribute the positive *MR* contribution to the suppression of superconducting fluctuations (i.e., the Aslamazov–Larkin (AL) contribution) by applying magnetic fields. Similar determination of the superconducting fluctuation has been reported for YBa$_2$Cu$_3$O$_{6+\delta}$ [18]. In this study for $x = 0.4$, the observed $T_{scf}$ ($T_{scf}$ = 40 K) was 2.7 times larger than $T_c$ ($T_c$ = 14.7 K). Similar to the case of $x = 0.4$, $T_{scf}$ was estimated at 28, 16, and 15 K for $x = 0.3$, 0.2, and 0.1, respectively. To visualize these onsets, the *MR* at 9 T is plotted as a function of the temperature in Fig. 3. These $T_{scf}$ values were cross-checked by the magnetic susceptibility measurements [19].

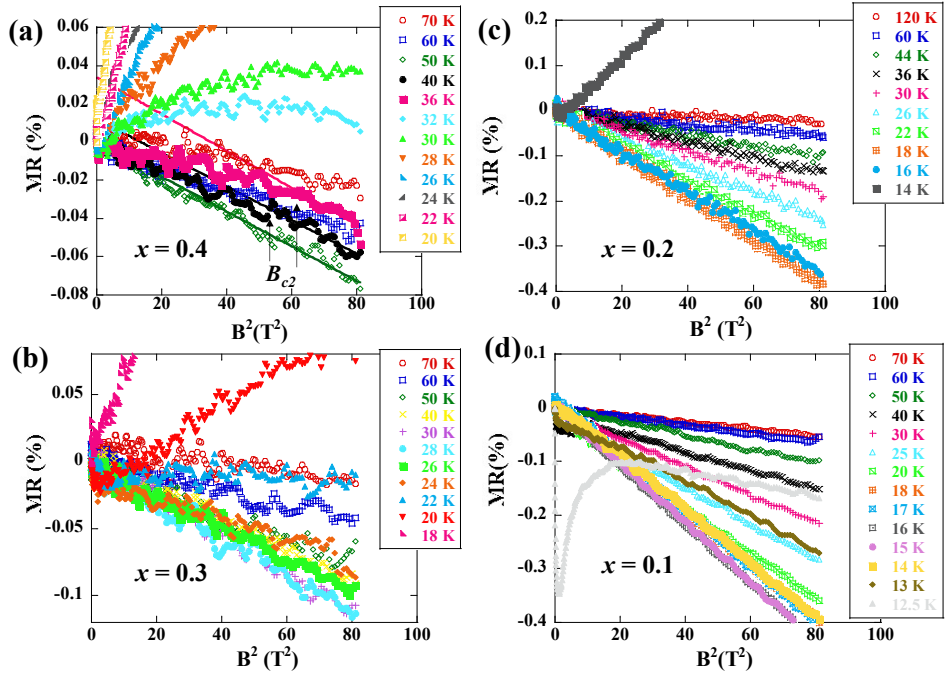

**Fig. 2.** (Color online) *MR* at different temperatures for the Te-annealed FeTe$_{1-x}$Se$_x$ single crystals with (a) $x = 0.4$, (b) $x = 0.3$, (c) $x = 0.2$, and (d) $x = 0.1$. (a) Normal state $B^2$-dependence of *MR* are drawn by straight lines for eye guides at 50, 40, and 36 K. $B_{c2}$ values are indicated by arrows.

It should be noted that qualitatively different behaviors have been reported for O$_2$-annealed FeTe$_{0.6}$Se$_{0.4}$ [20]. In ref. [20], *MR* is positive below 100 K, and it increases linearly with the applied field from an intermediate field (2 T at 16 K) to the measurement limit of 14 T. This anomalous field dependence has been interpreted by the existence of the Dirac cone state. However, we could not observe such *B*-linear *MR* in Te-annealed crystals. We suppose that band structures between O$_2$ and Te-annealed crystals are slightly different. Further studies are still needed to clarify the difference in those band structures.

## 4. Discussion

In the BCS-BEC crossover regime, the size of the pairs ($\xi_{ab}$) is comparable to the average inter-particle distance ($\sim k_F^{-1}$), i.e., $k_F\xi_{ab}$ is in the order of 1 [5]. Table I summarizes the parameters used in this study. Here, the in-plane coherence length $\xi_{ab}$ was estimated using $B_{c2}(0) = \Phi_0/2\pi\xi_{ab}^2$, where $\Phi_0$ is the flux quantum. The Fermi wave number $k_F$ (= 2.72 nm$^{-1}$) was estimated assuming the three-dimensional model as $k_F = (3\pi^2 n_h)^{1/3}$, because the anisotropy parameter $\gamma$ is reported to be small ($\gamma$ = 1.5 - 2) [21]. Here, $n_h$ is the hole carrier density obtained from our previous Hall coefficient $R_H$ measurements via $R_H = 1/en_h$ for an as-grown FeTe$_{0.8}$Se$_{0.2}$ single crystal (dominant carriers are holes) at 15 K [12], which was assumed to be common for all the samples in this study. Consequently, $k_F\xi_{ab}$ was estimated as approximately equal to 5, which is in the order of 1.

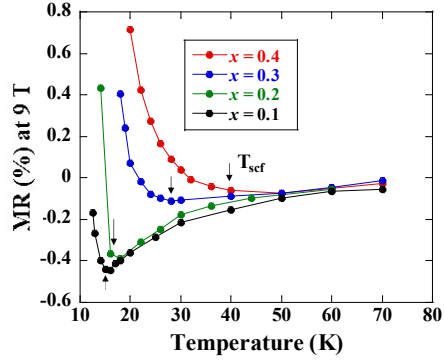

**Fig. 3.** (Color online) Temperature dependence of MR at 9 T for the Te-annealed FeTe$_{1-x}$Se$_x$ single crystals with $x$ = 0.4, 0.3, 0.2, and 0.1. The arrows indicate the onset temperatures for the superconducting fluctuations, $T_{scf}$.

Moreover, in this regime, the effect of superconducting fluctuations is expected to be enhanced because Cooper pairs can form at temperatures higher than $T_c$ [2]. For $x$ = 0.4, the $T_{scf}/T_c$ ratio was estimated at 2.7. This value is larger than that for FeSe (i.e., 2.2) [2] or slightly overdoped Bi-2212 (i.e., 1.3) [22].

**Table I.** Obtained parameters for the Te-annealed FeTe$_{1-x}$Se$_x$ single crystals with $x$ = 0.1, 0.2, 0.3, and 0.4. Here, $T_c$ is determined by half of normal state resistivity. The methods for estimating other parameters are described in the text.

| Se concentration $x$ | 0.1 | 0.2 | 0.3 | 0.4 |
|---|---|---|---|---|
| $B_{c2}(0)$ (T) | 81.1 | 86.6 | 103.3 | 68.2 |
| $\xi_{ab}$ (nm) | 2.02 | 1.95 | 1.79 | 2.20 |
| $k_F\xi_{ab}$ | 5.5 | 5.3 | 4.9 | 6.0 |
| $T_c$ (K) | 11.2 | 12.5 | 14.4 | 14.7 |
| $T_{scf}$ (K) | 15±1 | 16±2 | 28±2 | 40±2 |
| $T_{scf}/T_c$ | 1.3 | 1.4 | 1.9 | 2.7 |

## 5. Conclusion

The in-plane resistivity $\rho_{ab}$ of high-quality Te-annealed superconducting FeTe$_{1-x}$Se$_x$ ($x$ = 0.1, 0.2, 0.3, and 0.4) single crystals was examined under magnetic fields parallel to the $c$-axis. The resistive transition curves' analysis showed that $B_{c2}(0)$ was 60–100 T and $\xi_{ab}$ was approximately 2 nm. The short coherence length, $\xi_{ab}$, combined with a large inter-particle distance, $\sim k_F^{-1}$, were found to satisfy the criterion for being situated in the BCS-BEC crossover regime (i.e., $k_F\xi_{ab}$ was in the order of 1). Furthermore, *MR* measurements revealed that $T_{scf}$ was very high compared to $T_c$ (i.e., $T_{scf}$ = (1.3–2.7) $T_c$). These results provide further evidence for the occurrence of the BCS-BEC crossover in this system.

However, the data presented in Table I reveals that $T_{scf}/T_c$ decreased with decreasing Se concentration (i.e., $x$). The reason is remained as an open question.

The high $T_{scf}$ values demonstrated that the Te-annealing is greatly effective to prepare high-quality FeTe$_{1-x}$Se$_x$ single crystals. Owing to this novel annealing method, the FeTe$_{1-x}$Se$_x$ has become a suitable system for studying the behaviors regarding the BCS-BEC crossover. In this study, the proposed method was successfully applied in magneto transport measurements. It is further expected to be employed in various measurements such as magnetic susceptibility [19], ARPES, and scanning tunneling microscopy to clarify the crossover phenomena and unveil the rich physics of this fascinating material.

## Acknowledgment

This work was supported by JSPS KAKENHI Grant No. 20K03849, Iwate University, and Hirosaki University Grant for Distinguished Researchers from fiscal years 2017 to 2018.